\documentstyle[12pt]{article}
\renewcommand{\b}{\beta}
\newcommand{\g}{\gamma}

\newcommand{\e}{\epsilon}
\newcommand{\Qu}{{\rm Qu}}
\newcommand{\qu}{{\rm qu}}

\newcommand{\ar}{\longrightarrow}

\newcommand{\nn}{\medskip}

\newcommand{\w}{\omega}

\newcommand{\la}{\lambda}
\renewcommand{\a}{\alpha}
\begin{document}
\title{Speedup of iterated quantum search by parallel performance}
\author{Yuri Ozhigov
\thanks{Department of applied mathematics, Moscow state 
 university of technology "Stankin", Vadkovsky per. 3a, 101472, Moscow, Russia,
 e-mail: \ y@ oz.msk.ru}}
\date{}
\maketitle
\begin{abstract}
Given a sequence $f_1 (x_1 ), f_2 (x_1 , x_2 ), \ldots , f_k (x_1 ,
\ldots , x_k )$ of Boolean functions, each of which $f_i$ takes the value 1 
in a single point of the form $x_1^0 , x_2^0 , \ldots , x_i^0$, $i=1,2,
\ldots , k$. A length of all $x_i^0$ is $n$, $N=2^n$. 
It is shown how to find $x_k^0$ $\ (k\geq 2)$ using 
$\frac{k\pi\sqrt{N}}{4\sqrt{2}}$
simultaneous evaluations of functions of the form $f_i ,f_{i+1}$
with an error probability of order $k/\sqrt{N}$ which is $\sqrt{2}$ 
times as fast as by the $k$ sequential
applications of Grover algorithm for the quantum search.
Evolutions of amplitudes in parallel quantum computations are approximated by 
 systems of linear differential equations.
Some advantage of simultaneous evaluations of all $f_1 ,\ldots f_k$ are 
discussed.
\end{abstract}

\section{Introduction}

\subsection{Structure of the work}

After background and setting of a problem I present the short introduction 
 to abstract quantum computations in the section 3. In the section 4 linear
differential equations are applied to a tight analysis of the famous Grover
algorithm of the fast quantum search. 

The section 5 is the key. Here a parallel quantum algorithm for repeated search
 is defined and studied by means of differential equations. 
In the section 6 we briefly run through a parallel algorithm for iterated 
search with simultaneous queries of all oracles. 
The abstract includes the other substantiation of the algorithm.
  
\subsection{Background}

The most promising quantum mechanical application to the algorithm theory is 
associated with the fundamental algorithm of exhaustive search or finding a 
solution
of equation $f(x)=1$ for a Boolean function $f$. In 1996 Lov Grover 
in the work \cite{Gr1} showed how quantum computer can solve this equation for 
the case of unique solution 
in a time $O(\sqrt{N})$ where $N$ is the number of all possible values for 
$x$, 
whereas every probabilistic classical algorithm requires a time 
$O(N)$. 
At about the same time in the work \cite{BBBV} it was shown that
there are not substantially faster algorithm for this problem. Later 
a tight estimation for the time of Grover's algorithm as 
$\frac{\pi\sqrt{N}}{4}$ with the probability of error about $1/N$ 
was established in the work \cite{BBHT}.
A further development of the fast quantum search can be found in the works
\cite{DH}, \cite{FG}, \cite{Jo}, \cite{CGW}, \cite{Ro}, \cite{BBBGDL},
\cite{JMH}, \cite{H}.

  The earlier patterns of quantum speedup was constructed by P. Shor 
(look at the work \cite{Sh}). There are the algorithms finding a factorization 
of an integer and a discrete logarithm. Algorithms of such a sort are presented 
in a lot of works (look for example at \cite{DJ}, \cite{Gr2}, \cite{Ki1}, 
\cite{Si}, \cite{St}, \cite{TS}, etc.).

 The classical computations admitting quantum speedup are rare exclusions from 
all classical computations in the following sense. 
   Denote all words of a length $n$ in the alphabet $\{ 0,1\}$ by $\{ 0,1\}^n$.
We can represent a general form of classical computation as $T$ iterated 
applications of some oracle $g:\ \{ 0,1\}^n \ar \{ 0,1\}^n$:
\begin{equation}
x\ar g(x)\ar g(g(x))\ar\ldots\ar g(g(\ldots (g(x)))).
\label{1}
\end{equation}
In the work \cite{Oz1} it is shown that if 
$T=O(N^{\frac{1}{7+\e}} ),\ \e>0$, then for the bulk of all $g$ such 
computation has not any quantum speedup. Similar results for search 
problems were obtained 
in the works \cite{BBBV}, \cite{BBHT}, \cite{Oz2}, \cite{Za}. 
A lower bound as $O(N)$
for a time of quantum computations of functions with functional argument
$F:\ \{ f\}\ar\{ 0,1\}$ was found in the work \cite{BBCMW}. Here $f$ are 
Boolean 
functions on domain of cardinality $N$.
At the same time using a memory of $O(N)$ qubits it is possible to compute 
such functions in a time $N/2$ (look at \cite{vD}). This brings up the 
question: what a general type of classical computations of the form (1) 
admits a quantum speedup beyond any possible speedup of $g$? As follows 
from the work \cite{Oz1} for the bulk of functions $g$ this speedup can 
result only from a parallel application of $g$. 

  About other aspects of quantum evolutions look also at \cite{ML}, \cite{Ho},
\cite{Pe}.

\section{Setting of the problem}

Consider the following situation.  We want to gather a mosaic from scattered 
stones in a rectangular list with the corresponding picture. 
Each stone is of a unique form. We can gather this mosaic layer by layer and 
use a simple search among stones still scattered to fill any layer basing on 
the previous one. Then we in fact fulfill an iterated search classically, 
because to find the stones for the following layer we must already have the 
previous one filled. 

We formalize this as the special type of iterated algorithms: iterated search 
(IS).
Suppose we have a sequence $S_1 , S_2 ,\ldots , S_k$ of similar search 
problems where $S_i$ is to find a unique solution $x_i^0$ of equation 
$f_i (x_i )=1$ where a Boolean function $f_i$ is accessible iff we know 
all $x_j^0 ,\ \ j<i,$
$|x_i |=n,\ \ N=2^n ,\ \ k\ll N$, $|x|$ denotes a length of word $x$.
The aim is to discover $x_k^0$, $k\geq 2$. 
In view of the result of the work \cite{BBHT} 
cited above sequential applications of Grover's search for 
$x_1^0 ,x_2^0 ,\ldots , x_k^0$ give an answer in the time 
$\frac{k\pi\sqrt{N}}{4}$ with error probability about $k/N$. To do this we 
must have all oracles $f_i,\ \ i=1,2,
\ldots , k$, where the dependence $f_i$ of all $x_j ,\ \ j<i$ can be included 
to $f_i$. So we can assume that $f_i$ has the form $f_i 
(x_1 , x_2 ,\ldots ,x_i )$ and each equality $f_i (x_1 , \ldots , x_i )=1$ 
has the unique solution $x_1^0 ,
x_2^0 , \ldots , x_i^0 ,\ \ i=1,2,\ldots , k.$ Considering all oracles $f_i$ 
as 
 physical devices which can not be cloned we assume that they are in our 
disposal at the same time, so we can apply them simultaneously. 
In such application advantage is taken of interference between the results 
of their actions. This results in a speedup of computation comparatively with 
the sequential mode.
Why this speedup can arise? It arises because of a leak of amplitude in each 
step of sequential search. A leak of amplitude issues from that an amplitude 
of 
$x_i^0$ in search number $i$ increases step by step in course of Grover's 
search, after few first $l$ steps it becomes approximately
$\frac{2l+1}{\sqrt{N}}$, when amplitudes of others $x_i \neq x_i^0$ decrease.
This prevailing of the amplitude corresponding to $x_i^0$ 
(a leak of amplitude) can be immediately used for the next $i+1$-th search.
We shall show how this effect can be used to solve the problem of iterated 
search in a time $O(\frac{k\pi\sqrt{N}}{4\sqrt{2}})$ which is $\sqrt{2}$ 
times as fast as by $k$ sequential applications of Grover's algorithm.
Thus we shall have a modification of the fast quantum search - the parallel 
quantum algorithm for iterated search which will be described later. 
In this article we take up mostly the particular case $k=2$ of IS, we name 
this problem repeated search (RS).

RS-problem is connected with the known problem of structured search (SS). 
The problem of structured search is to find a unique solution $x_0 ,y_0$ of 
$f(x,y)=1$ provided we have a function $g$ whose support $\{ x\ |\ g(x)=1\}$ 
of cardinality $M$ contains $x_0$. RS-problem is a particular case of SS when 
$M=1$. The case $1\ll M\ll N$ was investigated by Farhi and Gutmann in 
the work \cite{FG}. They have found quantum algorithm for this case with a 
time complexity $O(\sqrt{MN})$, and also they wrote that the best known 
strategy for the case $M=1$ is the sequential application of Grover algorithm. 
In the present paper it is shown how this evident strategy can be improved 
in constant factor $\sqrt{2}$. At last note that our approach differs from 
the work \cite{FG}. Farhi and Gutmann used only algebraic properties of 
Grover's algorithm whereas for RS-problem we need to work with an evolution 
of amplitudes in computation.

\section{ Quantum computations and differential equations}

After early studies of R. Feynman (\cite{Fe}), P. Benioff (\cite{Be}) 
and D. Deutsch (\cite{De}) numerous approaches to quantum computations 
have appeared (look at \cite{BV}, \cite{Wa}, \cite{Ya}, \cite{Ll1}).
Leaving aside the problem of decoherence and quantum codes (look at the 
articles \cite{AB} , \cite{CLSZ}, \cite{Ki2}, \cite{Pr} ) we 
shall regard ideal computations in closed systems.
We use the abstract model of quantum computer independent of the formalism 
of classical algorithm theory. This model consists of two parts: classical 
and quantum (look at the picture 1). 

A state $C$ of {\it Classical part} consists of the following objects.

 1) Registers with labels corresponding to transformations $U_{i_j}$ 
of the finite set $\{ U_i \}$ of elementary unitary transformations with no 
more than 3 qubits each. (Strictly speaking transformations on two qubits 
would suffice: look for example at the work \cite{DiV}). 
Moreover, as follows from the works \cite{Ll2}, \cite{BBCDMSSSW} there is a 
variety of possible choices of the set $\{ U_{i_j} \}$.

2) Pointers aimed from these registers to as many qubits from the quantum part 
as there are arguments of the corresponding unitary transformation.
 Here each qubit is involved in exactly one transformation.

3) Registers of an end of computation and of a query: $e(C)$
and $\qu (C)$ respectively.

A {\it Quantum part} is a tape partitioned into cells with one qubit each. 
Every qubit has two basic states $|0\rangle ,\ |1\rangle$, so its quantum 
state
$\la |0\rangle +\mu |1\rangle$, $|\la |^2 +|\mu |^2 =1$, belongs to the
curcle of radius 1 in two dimensional Hilbert space ${\rm C}^2$. If $n$ is a 
length of tape, all states of the tape belong to the tensor product
${\cal H} =\underbrace{{\rm C}^2 \bigotimes {\rm C}^2 \bigotimes\ldots 
\bigotimes {\rm C}^2}
_{n}$ of spaces, corresponding to all qubits that is ${\cal H}={\rm C}^{2^n}$.
Each state of quantum part is a superposition $\chi =\sum\limits_{i=0}^{N-1} 
\la_i e_i$ of basic states $e_0 ,\ldots , e_{N-1}$ with complex amplitudes 
$\la_i$ where $\sum\limits_{i=0}^{N-1} |\la_i |^2 =1$, $N=2^n$.
We can assume that all $e_i \in\{ 0,1\}^n$.

An {\it Observation} of this state $\chi$ is a random variable which takes 
each value $e_i$ with the probability $|\la_i |^2$.

A {\it Working transformation} of quantum part corresponding to a fixed state 
of classical part has the form
$U_{i_1} \bigotimes U_{i_2} \bigotimes\ldots \bigotimes U_{i_k}$ where each 
$U_{i_j}$ acts on qubits which the corresponding pointer aims to.

Let $f_1 , \ldots , f_l$ be functions of the form $\{ 0,1\}^n \ar \{ 0,1\}^m$ 
and for each $i=1,2,\ldots , l$ there are special places in the quantum tape 
reserved for an argument $a_i$ of $f_i$ (query) and for a value of $f_i$ 
(answer).
Denote by $b_i$ an initial contents of the place for answer.

A {\it Query transformation} $\Qu _{\bar f}$ is $\Qu _{f_1} \bigotimes 
\Qu _{f_2}
\bigotimes\ldots \bigotimes \Qu _{f_l}$ where for each $i=1,\ldots , l$
\newline $\Qu _{f_i} \ |a_i ,b_i \rangle \ar |a_i , b_i \bigoplus f_i (a_i ) \rangle$, 
$
\bigoplus$ is bitwise addition modulo 2. We name these functions $\Qu_{f_i}$ 
oracles.

A {\it Quantum algorithm} is an algorithm determining evolution of the 
classical part:
\begin{equation}
C_0 \ar C_1 \ar \ldots \ar C_T
\label{2}
\end{equation} 
(in particular it determines a number $T$).
A classical part plays the role of controller for quantum part and determines 
its evolution (look at the picture 2).

A {\it Quantum computation} consists of two sequences: (2) and
\begin{equation}
Q_0 \ar  Q_1 \ar \ldots \ar Q_T
\label{3}
\end{equation}
where for each $i=0,1,\ldots ,T-1$ $e(C_i )=0;$
 $e(C_T )=1$ and every passage $Q_i \ar Q_{i+1}$ is:

- a working transformation, corresponding to $C_i$, if $\qu (C_i )=0,\ 
e(C_i )=0$,

- a query transformation $\Qu_{\bar f}$, if $\qu (C_i )=1,\ e(C_i )=0$.

A {\it result} of this quantum computation is a contents of first 
$n_0$ qubits of quantum tape after the observation of final state 
$Q_T$. An initial state $(C_0 , Q_0 )$ of the computer is obtained from an 
input data $\bar x$ by some routine procedure.

A {\it time} of computation (2),(3) is a number of query transformations 
(queries) in it.
We see that in this model some oracles may be called simultaneously which 
causes interference between results of their actions.
We shall prove that such interference can speed up computations in case of 
repeated search.

From a physical standpoint systems of linear differential equations is a natural 
tool for the study of quantum computation. Wave function $\psi$ which determins
an evolution of quantum computer satisfies Shr\"odinger equation
$\frac{\partial \psi}{\partial t} =iH\psi$, where $H$ is Hamiltonian within 
a real factor. An evolution of $\psi$ is determined by unitary operator
$U(t):$ $\psi (t)=U\psi (0)$ which satisfies the equation $\dot U=iHU$. This 
 equation is a prototype of all systems of differential equations studied below.
We need only to choose Hamiltonian so that the amplitude of target state peaks
 in a point $t_{quant}$ which is less than the time of the best classical 
computation. In rare cases quantum parallelism makes it possible.
 
Some other aspects of parallelism in computing may be found in 
\cite{BM}, \cite{BO}, \cite{LMS}, \cite{Wo}.

\section{ An exact description of simple quantum search by differential 
equations}

\subsection{Notations}

Assume the notations of Dirac where a vector $\bar a\in {\rm C}^m$ as a column 
of coordinates is denoted by $|\bar a \rangle$. A row obtained from $|\bar 
a\rangle$ by the transposition and complex conjugation is denoted by 
$\langle \bar a |$. A dot product of $\bar a,\bar b\in {\rm C}$ will be 
$\langle \bar a|\bar b\rangle$.
A result of application of operator $A$ to a vector $|\bar a\rangle$ is 
denoted by $A|\bar a \rangle$. For every transformations 
$A,\ B$ of the forms ${\cal L}_1 \ar {\cal L}_2 ,\ {\cal L}_2 \ar {\cal L}_3$ 
we denote by $AB$ their composition which acts from right to left such that 
$AB(x) 
=A(B(x))$.
Given vectors $a\in{\cal L} ,\ b\in {\cal M}$ from linear spaces $\cal L$, 
$\cal M$ the state $|a\rangle\bigotimes |b\rangle \in {\cal L}\bigotimes 
{\cal M}$ is denoted by $|a,b\rangle$. 
For a function $F:\ \ F|X,Y\rangle =|X,\phi (Y) \rangle$ we denote by 
$F|_Y$ its restriction on $Y:\ \ F|_Y |Y\rangle =|\phi(Y)\rangle$.

Let $f$ be a function of the form $A\ar A$. We define an $i$-th iteration of 
$f$: $f^{\{i\}}$ by the following induction on $i$.
Basis: $f^{\{1\}} =f$.
Step: $f^{\{i+1\}} =ff^{\{i\}}$.

\subsection{ Grover's quantum algorithm for simple search and its 
implementation in our model}

Grover's algorithm for finding a unique solution $x_0$ of equation 
$f(x)=1$ for a Boolean $f$ is sequential applications of the following 
unitary transformation: $G= -WR_0 WR_t$ to the initial state $\chi_0 =
\frac{1}{\sqrt{N}}\sum\limits_{i=0}^{N-1} |e_i \rangle$ where Walsh-Hadamard 
transformation is $W=\underbrace{J\bigotimes\ldots\bigotimes J}_{n}$, 
$$
J=\left(\begin{array}{cc} 1/\sqrt 2\ &1/\sqrt 2 \\
1/\sqrt 2 \ &-1/\sqrt 2
\end{array}\right),
\ \ R_0 |e\rangle = \left\{ \begin{array}{cc} |e\rangle ,\ &\mbox{if }  e\neq 
\bar 0,\\
-|0\rangle ,\ &\mbox{if } e=\bar 0,
\end{array}
\right.\ \ 
R_t |e\rangle = \left\{ \begin{array}{cc} |e\rangle ,\ &\mbox{if } 
e\neq x_0,\\
-|x_0 \rangle ,\ &\mbox{if } e=x_0 .\end{array} \right.
$$
It is easily seen that $W$ can be implemented on our model of quantum 
computer.

To implement $R_t$ it is sufficient to apply $\Qu_f$ to the state
$\frac{1}{\sqrt{N}} \sum\limits_{i=0}^{N-1} |e_i \rangle \bigotimes
\frac{|0\rangle -|1\rangle}{\sqrt{2}}$, where the last qubit is the place 
for oracle's answer.

To implement $R_0$ we need $n+1$ ancillary qubits initialized by $0$.
Consider some function $\phi$ acting on three qubits:
$|\mbox{main, ancilla, }res\rangle$ as follows.
$$
\begin{array}{cc}
|000\rangle\ &\ar\  |000\rangle\\
|100\rangle\ &\ar\ |101\rangle\\
|001\rangle\ &\ar\ |001\rangle\\
|101\rangle\ & \ar\ |111\rangle .
\end{array}
$$
Apply $\phi$ sequentially after each step moving pointers to right on
one qubit in the main and ancillary areas (look at the picture 3). 
This makes $res =1$ iff at least one of the main qubits is 1. Then inverse
the phase of $0$ in the qubit $res$ and fulfill all reverse 
transformations with $\phi$ 
 in the reverse order restoring the initial states of ancillary qubits.

Let $\chi_i =a_i \sum\limits_{e' \neq e} |e'\rangle +b_i |e\rangle ,$
$\chi_{i+1} =G \chi_i$, $e$ is a state of quantum part, corresponding 
to a target word $x_0$. The difference between $x_0$ and $e$ is that $e$ 
contains also ancillary qubits whose values will be restored after each 
step of computation (it can be simply traced in what follows). The main 
property of Grover's transformation may be represented by the following 
equations (look at \cite{Gr1}, \cite{BBHT} ).
\begin{equation}
\left\{\begin{array}{cl}
b_{i+1} &= (1-\frac{2}{N} )b_i +2(1-\frac{1}{N} )a_i ,\\
a_{i+1} &=-\frac{2}{N} b_i +(1-\frac{2}{N} )a_i .\\
\end{array}\right.
\label{4}
\end{equation}

\subsection{ The passage to the system of differential equations }

The system (4) can be rewritten in the following form
\begin{equation}
\left\{\begin{array}{cl}
b_{i+1} -b_i &= -\frac{2}{N} b_i +2(1-\frac{1}{N} )a_i ,\\
a_{i+1} -a_i &=-\frac{2}{N} b_i -\frac{2}{N} a_i ,\\
\end{array}\right.
\label{5}
\end{equation}
where the initial condition of quantum part gives 
$a_0 =b_0 =1/\sqrt{N}$.
This system (5) with the initial condition is the system of difference 
equations approximating the following system of linear differential 
equations:
\begin{equation}
\left\{\begin{array}{cl}
\dot b\delta &= -\frac{2}{N} b +2(1-\frac{1}{N} )a,\\
\dot a\delta &=-\frac{2}{N} b -\frac{2}{N} a,\\
\end{array}\right.
\label{6}
\end{equation}
with two unknown functions $a(t), b(t)$, constant $\delta >0$ and the 
initial condition $a(0)=b(0)=\/\sqrt{N}$, where $a_i ,b_i$ approximate 
$a(i\delta ), b(i\delta )$; $\delta$ is a step.
A difference between solutions of (5) and (6) on a segment of the form 
$t\in [0,O(\delta\sqrt{N} )]$ is $O(\sqrt{N} \delta ^2 )$, hence the 
error of this approximation may be done as small as required by varying 
$\delta$ (an integral part of number $x$ is denoted by $[x]$).

Solving (6) we obtain
$$
\ddot b +\frac{4}{\delta^2 N} b+O(\frac{b}{\delta^2 N^2} )+\ddot b O(
\frac{1}{N} )+\dot b O(\frac{1}{N\delta })=0.
$$
Hence in within
$O(\frac{1}{\sqrt{N}})$ a solution $b$ of (6) can be approximated by
a solution of equation
\begin{equation}
\ddot b +\frac{4}{\delta^2 N} b=0
\label{7}
\end{equation}
with the initial conditions $b(0)=\frac{1}{\sqrt{N}},$ $\frac{1}{2}
(\dot b (0)\delta +\frac{2}{N} b(0))=\frac{1}{\sqrt{N}}$ on the segment
$[0,2/\w ]$, where $\w = 2/\delta \sqrt{N}$.
The required solution of (7) with this accuracy is $b=\sin (\w t)+
\frac{1}{\sqrt{N}}\cos (\w t)$, and the maximum of amplitude 1 is in the point 
$t_0 = \frac{\pi\sqrt{N}}{4}\delta -\frac{\delta}{2}$. Then 
$\left[\frac{t_0}{\delta}\right] =\left[\frac{\pi\sqrt{N}}{4}\right]
\stackrel{+}{-} 1$ recurrent steps (4) are necessary and sufficient to achieve 
this value of $b$ with this accuracy 
Thus we obtain that the accuracy $O(\frac{1}{\sqrt{N}})$ is reached in 
$\left[\frac{\pi\sqrt{N}}{4}\right]$ Grover's iterations. In the work 
\cite{BBHT} its authors obtained an exact solution of (4) and thus proved 
that in fact a probability of error is even about $1/N$. The approximation 
of amplitude's evolution by systems of differential equations is more 
universal method. For example it makes possible to handle with the more 
involved case of parallel algorithm for RS which is the subject of next 
section.

\section{Parallel algorithm for the repeated quantum search}

\subsection{ Definitions and result}

Let $u,\ x,\ y$ be variables with values from three different copies of 
${\cal H}_0 ={\rm C}^N$, $a=a_1 \bigotimes a_2 \in {\rm C}^4$, where
$a_1 =a_2 =\frac{1}{\sqrt{2}} (|0\rangle -|1\rangle )$. 
We assume the notations $f_1 (x),\ f_2 (x,y)$ for two oracles in the repeated 
quantum search and let $e_1, e_2$ be such values for $x,y$ which represents  
unique solutions of equations $f_1 =1,\ f_2 =1$.
We denote the corresponding states of quantum tape by the same letters.

Put ${\cal H}={\cal H}_0 \bigotimes {\cal H}_0 \bigotimes {\cal H}_0 
\bigotimes {\rm C}^4$. 
Let
$$
\begin{array}{rl}
F_1 |u,x,y,a\rangle &=|u,x,y,a_1 \bigoplus f_1 (u),a_2 \rangle ,\\
F_2 |u,x,y,a\rangle &=|u,x,y,a_1 ,a_2 \bigoplus f_2 (x,y)\rangle ,\\
P|u,x,y,a\rangle &=|u\bigoplus x,x,y,a\rangle.
\end{array}
$$
Then 
$$
F_1 |u,x,y,a\rangle =\left\{
\begin{array}{cc}
|u,x,y,a\rangle ,\ &\mbox{if } u\neq e_1 ,\\
-|u,x,y,a\rangle ,\ &\mbox{if } u=e_1 ;
\end{array}
\right.
$$
$$
F_2 |u,x,y,a\rangle =\left\{
\begin{array}{cc}
|u,x,y,a\rangle ,\ &\mbox{if } |x,y\rangle\neq |e_1 ,e_2 \rangle ,\\
-|u,x,y,a\rangle ,\ &\mbox{if } |x,y\rangle =|e_1 ,e_2 \rangle ;
\end{array}
\right.
$$

Define the following auxiliary unitary transformations on $\cal H$:
${\cal R}_0 =I\bigotimes R_{0x} \bigotimes R_{0y} \bigotimes I;\ \ 
{\cal W} =I\bigotimes W_x \bigotimes W_y \bigotimes I;\ \ 
{\cal F} =P(F_1 \mid_{u,a_1} \bigotimes F_2 \mid_{x,y,a_2} )P,
$
where the lower indices $x,y$ point the corresponding area of application for 
Walsh-Hadamard transformations and rotations of the phase of 
$0$, $I$ denotes the identity.

The key unitary transformation of parallel algorithm for RS is
\begin{equation}
Z={\cal WR}_0 {\cal WF} .
\label{8}
\end{equation}

{\it The parallel algorithm for RS} is the sequential applications 
of $Z$ beginning with the initial state
$$
\chi_0 =| \bar 0\rangle \bigotimes\frac{1}{\sqrt{N}}\sum\limits_{i=0}^{N-1}
|e_i \rangle\bigotimes\frac{1}{\sqrt{N}}\sum\limits_{i=0}^{N-1}
|e_i \rangle \bigotimes a
$$
$\left[\frac{\pi\sqrt{N}}{2\sqrt{2}}\right]$ times.

\newtheorem{Theorem}{Theorem}
\begin{Theorem}
Let $t=\left[\frac{\pi\sqrt{N}}{2\sqrt{2}}\right]$. Then the observation
of $Z^{\{t\}} (\chi_0 )$ gives $x=e_1,\ y=e_2$ with probability of error
$O(\frac{1}{\sqrt{N}} )$.
\end{Theorem}

\subsection{ An advantage of parallel quantum algorithm}

It follows from the definition of $Z$ that oracles $F_1$ and $F_2$ for 
functions $f_1,\ f_2$ work simultaneously in parallel hence the algorithm 
requires approximately 
$\left[ \frac{\pi\sqrt{N}}{2\sqrt{2}} \right]$ simultaneous queries to obtain 
a result, when the sequential application of simple quantum searches with 
$f_1$ and then with $f_2$ requires $\left[ \frac{\pi\sqrt{N}}{2} \right]$ time 
steps to obtain a result with the same probability. Note that for a simple 
search 
constant factor $\frac{\pi\sqrt{N}}{4}$ can not be essentially improved (look 
at \cite{BBHT}).

Suppose that every query is fulfilled by a physical device (oracle) of the
peculiar type corresponding to a form of query.
A set with a minimum of oracles which is necessary for the solution of RS 
 consists of one oracle for $f_1$ and one for $f_2$. With these oracles we 
can run them simultaneously in the parallel algorithm and obtain a result 
$\sqrt{2}$ times faster than by the sequential search. But if we have {\it 
two copies} of each oracle it is possible to achieve the same performance 
by sequential search if we divide the whole domain $\{ 0,1\}^n$ into two 
equal parts of $N/2$ elements each and apply a simple quantum search at 
first with two copies of $f_1$-oracle - one for each part, then, having
$e_1$, with two copies of $f_2$-oracle. But this last way is expensive if 
every copy of oracle has a large cost, or impossible at all if every 
oracle is unique, say issues from a natural phenomenon. Just in this case 
of minimal possible set of oracles $f_1$, $f_2$ the application of 
parallel
quantum algorithm for RS gives the increasing of performance in $\sqrt{2}$ 
times. 
This speedup can be also obtained for IS if we apply this algorithm 
sequentially for the pairs 
$f_i , f_{i+1} ,\ i=1,2,\ldots , k-1.$ The resulting error probability will 
be $O(k/\sqrt{N})$.
The remaining part of this work is devoted to the proof of Theorem and 
perspectives of this approach.

\subsection{ A primary analysis of parallel algorithm for RS}

Note that each of $W_y ,\ R_{0y}$ commutes with $W_x ,\ R_{0x} ,\ P, \ F_1$;
$P$ commutes with $F_2$, hence $Z$ can be represented in the form
$$
Z=-[(I\bigotimes W_x R_{0x} W_x \bigotimes I)PF_1 P][-(I\bigotimes (
W_y R_{0y} W_y )\bigotimes I) F_2 ],
$$
or in the form
\begin{equation}
Z=\{ -W_x R_{0x} W_x {\cal F}_1 \}\{ -W_y R_{0y} W_y F_2 \} ,
\label{9}
\end{equation}
where 
$$
{\cal F}_1 |u,x,y,a\rangle =\left\{
\begin{array}{cl}
|u,x,y,a\rangle \ \ &\mbox{if } x\neq e_1 ,\\
-|u,x,y,a\rangle \ \ &\mbox{if } x=e_1 .
\end{array}
\right.
$$
The form (9) is exactly the repetition of Grover's transformations with 
oracles
$F_2 ,\ {\cal F}_1$ in this order, hence we can apply the formulas (4) for 
the resulting amplitudes of these transformations. 
Let the $Z$-iterations be
$\chi_0 \ar \chi_1 \ar\ldots\ar\chi_t$,
$\chi_{i+1} = Z(\chi_i ) ,\ i=0,1,\ldots , t-1$;

\begin{equation}
\chi_i = b_i |e_1 e_2 \rangle +a_i |e_1 N_2 \rangle +\a_i |N_1 N_2 \rangle
+\b_i |N_1 e_2 \rangle ,
\label{10}
\end{equation}
where $e_1$ and $e_1 ,e_2$ are the target states: unique solutions for
$f_1 (x)=1$ and for $f_2 (x,y)=1$ respectively, $N_1 =\sum\limits_{i=2}^N
e_i ,$ $N_2 =\sum\limits_{i\neq 2} e_i $ (we omit ancillary qubits).

We represent the transformation $\chi_i \ar Z(\chi_i )$ as two sequential 
steps:
$$
\chi_i \stackrel{1}{\ar} Z_1 (\chi_i )=\chi '_i \stackrel{2}{\ar} Z_2
(\chi '_i )=\chi_{i+1} ,
$$
where $Z_1 =-W_y R_{0y} W_y F_2 ,$ $Z_2 = -W_x R_{0x} W_x {\cal F}_1 .$
To calculate the change of amplitude resulting from the application of $Z_1$
(or $Z_2$) we shall fix a value of $x$ (or $y$ respectively).

{\bf Step 1}. Denote amplitudes of basic states in $\chi '_i$ by the 
corresponding letters with primes:
$$
\chi '_i = b'_i |e_1 e_2 \rangle +a'_i |e_1 N_2 \rangle +\a '_i |N_1 N_2 
\rangle +\b '_i |N_1 e_2 \rangle .
$$
Then for the two essentially different ways to fix a basic state
for $x$: $x=e_1$ or $x+e_j ,\ j\neq 1$ we shall have the different 
expressions
for new amplitudes. Use the property of the diffusion transformation 
$WR_0 W$ to be an inversion about average (look at \cite{Gr1}). Let $\la_{av}$ 
be an average amplitude of corresponding quantum state.

a). \underline{$x=e_1$. }

$\la_{av} = \frac{(N-1)a_i -b_i}{N} ,$ $b'_i =2\la_{av} +b_i$,
$a'_i =2\la_{av} -a_i$,
$$
\begin{array}{cl}
b'_i =\frac{2(N-1)a_i -2b_i}{N} +b_i &=b_i(1-\frac{2}{N} )+2a_i (1-
\frac{1}{N} ),\\
a'_i =\frac{2(N-1)a_i -2b_i}{N} -a_i &=-b_i \frac{2}{N} )+a_i (1-
\frac{2}{N} ).
\end{array}
$$

b). \underline{$x=e_j ,\ j\neq 1$. }

$\la_{av} =\frac{(N-1)\a_i +\b_i}{N} ,$ $\a '_i =2\la_{av} -\a_i$,
$\b '_i =2\la_{av} -\b_i$,
$$
\begin{array}{cl}
\a '_i =\frac{2(N-1)\a_i +2b_i}{N} -a_i &=\a_i (1-\frac{2}{N} )
+2\b_i \frac{2}{N} ),\\
\b '_i =\frac{2(N-1)\a_i +2\b_i}{N} -\b_i &=\a_i (1-\frac{1}{N} )-\b_i 
(1-\frac{2}{N} ).
\end{array}
$$

{\bf Step 2}. $\chi '_i \stackrel{2}{\ar} Z_1
(\chi '_i )=\chi_{i+1} .$

We have two different ways to fix a basic state for $y:$ $y=e_2$ or $
y=e_j ,\ j\neq 2$.

a). \underline{$y=e_2$. }

$$
\la_{av} =\frac{(N-1)\b '_i -b'_i}{N},
\begin{array}{cl}
b_{i+1} &= 2\la_{av} +b'_i =b'_i (1-\frac{2}{N} )+2\b '_i (1-\frac{1}{N} ),\\
\b_{i+1} &=2\la_{av} -\b '_i =\b '_i (1-\frac{2}{N} )-b'_i \frac{2}{N}.
\end{array}
$$

b). \underline{$y=e_j ,\ j\neq 2$.}

$$
\la_{av} =\frac{(N-1)\a '_i -a'_i}{N},
\begin{array}{cc}
a_{i+1} &= 2\la_{av} +a'_i =a'_i (1-\frac{2}{N} )+2\a '_i (1-\frac{1}{N} ),\\
\a_{i+1} &=2\la_{av} -\a '_i =\a '_i (1-\frac{2}{N} )-a'_i \frac{2}{N}.
\end{array}
$$

Hence, the recurrent formulas for amplitudes of sequential steps 1,2 acquire 
the following form:

$$
\begin{array}{cl} 
b_{i+1} &= b_i (1-\frac{2}{N} )^2 +2a_i (1-\frac{1}{N} )(1-\frac{2}{N} )+
4\a_i (1-\frac{1}{N} )^2 -2\b_i (1-\frac{2}{N} )(1-\frac{1}{N});\\
a_{i+1} &=a_i (1-\frac{2}{N} )^2 -b_i \frac{2}{N} (1-\frac{2}{N} )+2
\a_i (1-\frac{2}{N})(1-\frac{1}{N}) +2\b_i \frac{2}{N} (1-\frac{1}{N});\\
\a_{i+1} &=\a_i (1-\frac{2}{N})^2 +\b_i \frac{2}{N} (1-\frac{2}{N})-a_i 
(1-\frac{2}{N})\frac{2}{N} +b_i \frac{4}{N^2}; \\
\b_{i+1} &=2\a_i (1-\frac{1}{N})(1-\frac{2}{N}) -\b_i (1-\frac{2}{N})^2
-b_i (1-\frac{2}{N})\frac{2}{N} -2a_i (1-\frac{1}{N})\frac{2}{N} .
\end{array}
$$
Thus the matrix of one step of algorithm has the form
$$
Z=\left(
\begin{array}{cccc}
1 &2 &4 &-2\\
-\frac{2}{N} &1 &2 &\frac{4}{N}\\
\frac{4}{N^2} &-\frac{2}{N} &1 &\frac{2}{N} \\
-\frac{2}{N} &\frac{4}{N} &2 &-1
\end{array}
\right) .
$$

The system of recurrent equations can be rewritten as the following system of difference equations.
\begin{equation}
\begin{array}{cl}
b_{i+1} -b_i &=2a_i +4\a_i -2\b_i +b_i O_1 (\frac{1}{N}) +a_i O_2 
(\frac{1}{N})+\a_i O_3 (\frac{1}{N})+\b_i O_4 (\frac{1}{N});\\
a_{i+1} -a_i &= -\frac{2}{N} b_i +2\a_i +a_i O_5 (\frac{1}{N}) +
b_i O_6 (\frac{1}{N^2}) +\a_i O_7 (\frac{1}{N}) +\b_i O_8 (\frac{1}{N});\\
\a_{i+1} -\a_i &=-\frac{2}{N} a_i +\b_i O_{13}(\frac{1}{N}) +a_i O_{14}
(\frac{1}{N^2})+b_i O_{15}(\frac{1}{N^2})+\a_i O_{16}(\frac{1}{N}) ;\\
\b_{i+1} -\b_i &=-\frac{2}{N} b_i +2\a_i -2\b_i +a_i O_9 (\frac{1}{N}) +
\a_i O_{10} (\frac{1}{N})+\b_i O_{11}(\frac{1}{N}) +b_i O_{12}
(\frac{1}{N^2}) .
\end{array}
\label{11}
\end{equation}

\subsection{An approximation of amplitude's evolution by differential 
equations}

Let $\{\bar c_i \}$ be a sequence of vectors from ${\rm C}^k :$
$\bar c_i =(c_i^1 ,c_i^2 ,\ldots ,c_i^k ),$ $c_i^j \in \rm C$,
which satisfies the following system of difference equations
\begin{equation}
\bar c_{i+1} -\bar c_i =A\bar c_i,
\label{12}
\end{equation}
where $A$ is a matrix of size $k\times k$ with complex elements.

Let $m$ be an integer and a function $C(t):\ {\rm R}\ar{\rm C}^k$ is a
   solution of the system of differential equations 
\begin{equation}
\dot C(t) =mAC(t)
\label{13}
\end{equation}
with the initial condition 
\begin{equation}
C(0)=\bar c_0 .
\label{14}
\end{equation}
Then the exact solution of the Cauchy problem (13),(14) is 
$C(t)=R(t)\bar c_0 ,$ where the resolvent matrix $R(t) =\exp (mAt)$.
The system (12) will be the system of difference equations 
approximating $C(t)$ by Euler's method if we consider $\bar c_i$ as an
approximation of $C(i/m),\ i=0,1,\ldots .$
The accuracy of approximation may be obtained by the Taylor formula 
$C(\frac{i+1}{m})=C(\frac{i}{m})+\frac{1}{m} \dot C(\frac{i}{m})+
\frac{1}{2m^2}\ddot C(t_1 ),\ \frac{i}{m}<t_1 <\frac{i+1}{m}$.
Here the error $\e_1$ of the one step of the recursion (12) is the third 
summand
$\frac{1}{2m^2} \ddot C(t_1 )=\frac{1}{2} A^2 C(t_1 )$. Thus the error at the
first step is $\frac{1}{2} A^2 \exp (mA\theta_1 ) \bar c_0$,
at the second step: $\frac{1}{2} A^2 \exp (mA\theta_2 ) \bar c_1 +
\exp (mA\frac{1}{m}) \frac{1}{2} A^2 \exp (mA\theta_1 )\bar c0$
$= \frac{1}{2} A^2 \exp (mA\theta_2 ) (\bar c_0 +A\bar c_0 +\frac{1}{2}
A^2 \exp (mA\theta_1 )\bar c_0 )+\exp (A) \frac{1}{2} A^2 \exp (mA\theta_1 )
\bar c_0$, etc., at the $i$-th step the error will be
$\e_i \leq\frac{3}{2}\sum\limits_{j=1}{i} \exp (A\a_j )A^2 \bar c_0$,
where $0<\a_j <1$. Hence if $\| \bar c_0 \| \leq h$, then the error after 
$i$-th 
step is $\e_i =O(ih)$. Particularly, for the initial conditions
$\|c_0 \| =O(\frac{1}{N} )$ a good approximation can be obtained
if $i=o(N)$, and thus we can solve the Cauchy problem instead
of (11) for $i=O(\sqrt{N} )$ having error as small as required for sufficiently large $N$.

Define a new function $B(\tau )$ as follows 
$B (t m )=C(t)$. 
In terms of $B$ the Cauchi problem (13), (14) acquires the form 
\begin{equation}
\frac{{\rm d}}{{\rm d}\tau} B(\tau)=AB(\tau ),\ B(0)=c_0 .
\label{15}
\end{equation}
Apply this to the solution 
$\bar c_i =|b_i ,a_i ,\a_i ,\b_i \rangle$ of (11), where $\bar c_0 =
|\frac{1}{N} ,\frac{1}{N} ,\frac{1}{N} ,\frac{1}{N}\rangle$.
Put $B=|b,a,\a ,\b \rangle$ for the scalar functions $b,a,\b ,\a$ and
 denote the argument of the function $B$ by $t$. Then the equation
(15) approximating (11) acquires the following form:
\begin{equation}
\begin{array}{cl}
\dot b &=2a+4\a -2\b +bO_1 (\frac{1}{N} )+\e_1 +aO_0 (\frac{1}{N});\\
\dot a &=-\frac{2}{N} b+2\a +\e_2 +O_2 (\frac{1}{N})a;\\
\dot \b &=-\frac{2}{N} b+2\a -2\b +\e_4 +O_4 (\frac{1}{N}) a;\\
\dot \a &=-\frac{2}{N} a +\e_3,
\end{array}
\label{16}
\end{equation}
where $\e_i =a O_{0i} (\frac{1}{N^2}) +bO_{1i} (\frac{1}{N^2})
+\b O_{2i}(\frac{1}{N}) +\a O_{3i}(\frac{1}{N}),\ i=1,2,3,4,$
with the initial condition
\begin{equation}
b(0)=a(0)=\b (0)=\a (0)=\frac{1}{N} .
\label{17}
\end{equation}
Then for $t=O(\sqrt{N}),\ i=[t]$ the vector of error will be
$\bar\delta =\bar B(t) -\bar c_i =O(1/\sqrt{N}),\ \ N\ar\infty$ and with this 
accuracy we can write $b(i)\approx b_i$ for the amplitude $b_i$ of target 
state 
$|e_1 ,e_2 \rangle$.

\subsection{Tight analysis of the parallel quantum algorithm for RS}

Now we shall take up the system of linear differential equations (16) with the 
initial conditions (17). 
Our goal is to solve it on a segment of the form $0\leq t\leq O(\sqrt{N}).$
The system (16) can be represented in the form $\dot B =MB$, where its matrix 
$M=Z-1=\tilde A_0 +E+H$ ($1$ denotes the identity matrix) for the matrices
$$
\tilde A_0 =\left(
\begin{array}{cccc}
0 &2 &4 &0\\
-\frac{2}{N} &0 &2 &0\\
0 &-\frac{2}{N} &0 &0 \\
0 &0 &0 &0
\end{array}
\right),
E=\left(
\begin{array}{cccc}
0 &0 &0 &0\\
0 &0 &0 &0\\
0 &0 &0 &0 \\
-\frac{2}{N} &0 &2 &-2
\end{array}
\right),
H=\left(
\begin{array}{cccc}
 d_1 &d_1 &d_1 &-2+d_1\\
 d_2 &d_1 &d_1 &d_1\\
 d_2 &d_2 &d_1 &d_1 \\
 d_2 &d_1 &d_1 &d_1
\end{array}
\right),
$$
where $d_l$ denotes different expressions of the form $O(N^{-l} ),\ 
l=1,2.$ 

We shall show that the deposit of 
$\tilde A_0$ to the solution of (16),(17) is 
the main and deposits of $E$ and $H$ are negligible. What is the main 
difficulty here? Consider the resolvent matrix for the Cauchy problem (16),
(17), this is the solution $R(t)$ of the differential equation for matrices: 
$\dot R=MR$ with the initial condition $R(0)=1$. Then we have $C(t)=RC(0)$. 
The matrix $R$ has the form $\exp (Mt)$.
But in our case the matrices $\tilde A_0 ,E,H$ do not commutate, hence we cannot 
use the standard properties of exponent. In order to cope with this task 
we shall at first solve the Cauchy problem at hand neglecting deposits of $E$ 
and $H$ to the main matrix $M$. The legality of this approximation is
shown in the Appendix.

Now take up the reduced equation $\dot C(t) = AC(t)$ with the initial  
condition $C(0)=c_0$. Excluding the last column and row containing only 
zeroes we obtain the new matrix $A_0$. The characteristic equation for $A_0$ 
is $\la^3 
+\frac{8}{N}\la -\frac{16}{N^2}=0$ and its nonzero solutions in within
$O(\frac{1}{N})$ are $\la_{1,2} =\stackrel{+}{-} 2\sqrt{2}i/\sqrt{N}$.
Then standard calculations give the approximation of solution as
\begin{equation}
\begin{array}{cl}
b &=\frac{1}{2} -\frac{1}{2}\cos \frac{2\sqrt{2}t}{\sqrt{N}},\\
a &=\frac{1}{\sqrt{2N}}\sin \frac{2\sqrt{2}t}{\sqrt{N}}+\frac{1}{N} 
\cos\frac{2\sqrt{2}t}{\sqrt{N}},\\
\a &=\frac{1}{2N} \cos\frac{2t}{\sqrt{N}}+\frac{1}{2N}
\end{array}
\label{18}
\end{equation}
in within $|O(\frac{1}{\sqrt{N}}), O(\frac{1}{N} ), O(\frac{1}{N\sqrt{N}})
\rangle$. 
The amplitude $b$ from (18) peaks in the point $t_1 
=\frac{\pi\sqrt{N}}{2\sqrt{2}}$
where $b(t_1 )=1$ in within $O(\frac{1}{\sqrt{N}})$.
Assuming that deposits of $E$ and $H$ to the solution are  small, we obtain 
that the amplitude of target state $|e_1 ,e_2 \rangle$ will be 
$1-O(1/\sqrt{N} )$ after 
$\left[\frac{\pi\sqrt{N}}{2\sqrt{2}} \right]$ steps of parallel algorithm 
which is $\sqrt{2}$ times as small as the time of sequential quantum search. 

\subsection{Completion of the proof}

In the first version of this paper the deposit of $E$ and $H$ was estimated by the conventional procedure of approximation a solution of differential equation
(look at the Appendix). This is the immediate but cumbersome way to prove that this deposit is vanishing. After the publication of the first version of this paper Farhi and Gutmann informed me how to simplify this construction by uniting by pairs the sequential transformations in parallel algorithm (\cite{FG1}). In this section I combine this idea with the approach of the first version. 

At first turn to the orthonormal basis $E_1 = |e_1 e_2 \rangle$,
$E_2 =\frac{1}{\sqrt{N}}|e_1 N_2 \rangle$, $E_3 =\frac{1}{N}|N_1 N_2 \rangle$, $E_4 =\frac{1}{\sqrt{N}}|N_1 e_2 \rangle$. The matrix $Z$ in this basis acquires the form
$$
A_1 =\left(
\begin{array}{cccc}
1 &\frac{2}{\sqrt{N}} &\frac{4}{N} &-\frac{2}{\sqrt{N}}\\
-\frac{2}{\sqrt{N}} &1 &\frac{2}{\sqrt{N}} &\frac{4}{N}\\
\frac{4}{N} &-\frac{2}{\sqrt{N}} &1 &\frac{2}{\sqrt{N}} \\
-\frac{2}{\sqrt{N}} &\frac{4}{N} &\frac{2}{\sqrt{N}} &-1
\end{array}
\right) .
$$
Now group together each pair of unitary transformations in the algorithm:
$\chi_{2k} \ar\chi_{2k+1} \ar\chi_{2k+2}$ and denote by $B$ the corresponding matrix: $B_0 =A_1^2$. It is sufficient to prove that $\| B_0^{[\frac{\pi\sqrt{N}}{4\sqrt{2}}]}|0,0,1,0\rangle -|1,0,0,0\rangle\|
 =O(\frac{1}{\sqrt{N}})$, because one application of $A_1$ can only increase the 
error by $O(\frac{1}{\sqrt{N}})$. 

The Cauchy problem for the recursion 
$\bar c_{i+1} = B_0 \bar c_i$ has the form $\dot{\bar c} =(B_0 -1)\bar c,$ 
$\bar c=\bar c_0$, and its resolvent has the form $R=\exp Bt$, where $B=B_0 -1$. We have:
$$
B \approx\frac{4}{\sqrt{N}}
\left(
\begin{array}{cccc}
0 &1 &0 &0 \\
-1 &0 &1 &0 \\
0 &-1 &0 &0 \\
0 &0 &0 &0
\end{array}
\right) 
$$
in within $O(\frac{1}{N})$. 

Thus we can consider only a projection of $\bar c$ to the subspace 
${\cal H}_1$ spanned by 
$E_1 ,E_2 ,E_3$. Denote by $D$ the matrix 
$$
\left(
\begin{array}{ccc}
0 &-\frac{i}{\sqrt{2}} &0\\
\frac{i}{\sqrt{2}} &0 &-\frac{i}{\sqrt{2}}\\
0 &\frac{i}{\sqrt{2}} &0 
\end{array}
\right) .
$$
 Then the restriction of $B$ to ${\cal H}_1$ has the form 
$\frac{4\sqrt{2}i}{\sqrt{N}}D$. It is easily seen that 
$D^{2k+1}=D$, $D^{2k} =D^2$ for $k=1,2,\ldots$. 
If the number of steps is 
$\left[\frac{\pi\sqrt{N}}{2\sqrt{2}}\right]$ then
$t=\left[\frac{\pi\sqrt{N}}{4\sqrt{2}}\right]$. Here in within $O(\frac{1}{\sqrt{N}})$ we have
$$
\begin{array}{cc}
|b,a,\a\rangle &\approx \exp(\frac{4\sqrt{2}i}{\sqrt{N}} Dt)=
\exp (\pi i D)= \cos (\pi D)+i\sin(\pi D)\\
&=1-\frac{(\pi D)^2}{2}+\frac{(\pi D)^4}{4!}-\ldots +i(\pi D-
\frac{(\pi D)^3}{3!}+\ldots )\\
&=1-D^2 (1-\cos\pi )+iD\sin\pi =1-2D^2
\end{array}
$$
that is 
$$
\left(
\begin{array}{ccc}
0 & 0 & 1\\
0 & -1 & 0\\
1 & 0 & 0
\end{array}
\right) .
$$
The initial state is $|0,0,1\rangle$ in within $O(\frac{1}{\sqrt{N}})$. 
Application of this matrix to the initial state gives $|1,0,0\rangle$ 
with this accuracy. Theorem is proved.

\section{Parallel implementation of iterated quantum search}

\subsection{Parallel quantum algorithm for IS}
 
Now take up an IS problem for arbitrary $k$.
Consider an evolution of amplitudes arising when $k$ oracles work in parallel. 
Let $\chi_i = a_0^i |N_1 ,N_2 ,\ldots ,N_k \rangle +
a_1^i |e_1 , N_2 ,\ldots ,N_k \rangle +\ldots +a_k^i | e_1 ,e_2 ,\ldots ,
e_k \rangle +R_i$ (it generalizes (10)), where 
$R_i$ contains only basic states of the form \newline
$|\ldots ,N_p ,
\ldots ,e_q ,\ldots \rangle$.
The natural generalization of the transformation (8) will be
$$
Z_k =(-1)^k {\cal W}^{(k)} {\cal R}_0^{(k)}{\cal W}^{(k)} {\cal F}^{(k)},
$$
where
${\cal W}^{(k)} = W_1 \bigotimes W_2 \bigotimes \ldots
\bigotimes W_k \bigotimes
I$, each W-H transformation $W_i$ acts on $x_i$, $i=1,2,$ $\ldots ,$ $ k$,
${\cal R}_0^{(k)}= R_{01} \bigotimes \ldots \bigotimes R_{0k} \bigotimes I$,
each rotation of 0's phase $R_{0i}$ acts on $x_i$, $i=1,2,\ldots ,k$,
${\cal F}^{(k)}= F_1 \bigotimes\ldots \bigotimes F_k \bigotimes I$,
each $F_i$ acts on $x_i$ and inverses the sign of $e_i$, identities
$I$ act on ancilla. 

Let a matrix $A$ determines an evolution of quantum state in parallel 
algorithm such that $\chi_i =A \chi_{i-1}$. $A$ represents the operator in 
$2^{kn}$-dimensional space. We reduce $A$ to an operator $A_r$ acting on 
$k+1$-dimensional space generated by the vectors 
\newline $|N_1 ,N_2 ,
\ldots ,N_k \rangle , |e_1 , N_2 ,\ldots ,N_k \rangle ,\ldots ,
| e_1 , e_2 ,\ldots , e_k \rangle$. 
Then represent $A_r$ as $A_r =A_0 +B$, where $A_0$ is Jacobi matrix of the 
form
\begin{equation}
\left(
\begin{array}{ccccc}
0 &2 &0 &\ldots &0\\
-\frac{2}{N} &0 &2 &\ldots &0\\
\ldots &\ldots &\ldots &\ddots &\ldots\\
0 &\ldots &-\frac{2}{N} &0 &2\\
0 &\ldots &0 &-\frac{2}{N} &0\\
\end{array}
\right) .
\label{27}
\end{equation}

This matrix has the following nonzero elements: 2 - above the main diagonal 
and
$-\frac{2}{N}$ - behind it.
Its size is $(k+1)\times (k+1)$.
Assume that the effect of reduction: $A$ to $A_r$ and the deposit of 
$B$ are negligible. Generally speaking for $k>2$ this assumption should be 
proved. An evolution of amplitude can be represented approximately as the 
solution of Cauchy problem
\begin{equation}
\dot{\bar a} = A\bar a ,\ \ a(0) =|N^{-k/2} ,\ldots ,N^{-k/2} \rangle ,
\label{28}
\end{equation}
where $\bar a (i) \approx |a_1^i ,\ldots , a_k^i \rangle$, $i$ integer,
we assume that $k\ll \sqrt{N}$.

Given eigenvalues of matrix (19), a general solution can be obtained by the 
standard procedure. These eigenvalues for Jacobi matrix is known (look at 
\cite{Bel}, Chapter 2, ex. 32), there are
$\la_m =-\frac{4i}{\sqrt{N}} \cos m\theta ,\ \ m=1,2,\ldots ,k, \theta
=\frac{\pi}{k+2}$. We shall not solve (20) here.

\subsection{Perspectives of parallel quantum algorithm for iterated search}

One can ask: can we obtain a speedup by a big constant factor for iterated 
quantum search when applying parallel action of more than two oracles? 
In all probability the answer is no. 

Estimate 
the growth of target amplitude from above. 
Canceling all $-\frac{2}{N}$ we can only increase this growth.
This results in a simple system of linear differential equations whose 
solution is $a_k (t)=N^{-k/2} +2^k {\int\limits_0^t}^{\{ k\}} 
N^{-k/2} dt = N^{-k/2} + 2^k N^{-k/2} \frac{t^k}{k!}$. 
The parallel algorithm for 
IQS can exceed sequential quantum search only if $a_k (t)$ is substantially 
large for $t < k\sqrt{N}$. For example let the parallel algorithm work only 
the quarter of time required for the sequential search. Then it can not reach 
the vanishing error probability for any $k$, because for such 
$t=\frac{\pi\sqrt{N}k}{16}$ $a_k \approx \left(\frac{\pi}{8}\right)^k
\frac{k^k}{k!} <1$.

Nevertheless, parallel quantum algorithm has one advantage.
Compare sequential and parallel quantum algorithms for IS in case 
$1\ll k\ll\sqrt{N}$, if total time $t=\sqrt{N}$.
If $t=\sqrt{N}$ then $t$ sequential applications of $Z^{(k)}$ raise the
 amplitude $a_k$ up to the value $a_k (t)=\frac{a^k}{k!} ,\ \ a=2-\e$, where 
$\e\ar 0 \ \ (N\ar\infty)$. Hence the resulting probability is $P_{par} =
\left(\frac{a^k}{k!}\right)^2$. 

On the other hand if we have a total time $\sqrt{N}$ then we can apply 
sequential quantum searches with the time $\sqrt{N}/k$ for each $x_i^0 ,
\ \ i=1,2,\ldots ,k$. The probability $P_{seq}$ to find $x_k^0$ will be 
less than 
$\left(\frac{2}{k}\right)^{2k}$ because for one search it does not exceed
$\left(\frac{2}{k}\right)^2$. Consequently, $P_{par}$ exceeds $P_{seq}$ 
in more than $2^{2k}$ times.

\subsection{Conclusion}

To sum up, the parallel algorithm constructed above for repeated quantum 
search is $\sqrt{2}$ times as fast as sequential application of fast quantum 
search and it requires the same hardware. The advance is taken of interference 
arising when two oracles act simultaneously on the set of entangled qubits. 
This parallel quantum algorithm can be applied to the problem of $k$ dependent 
iterations of quantum search in areas of $N$ elements each, with the same 
effect of speedup in $\sqrt{2}$ times. 
Here the error probability will be vanishing if 
 $k=o(\sqrt{N}),\ N\ar\infty$. 

In addition, for the fixed total time $\sqrt{N}$ a probability of success 
for parallel algorithm is $2^{2k}$ times as big as for sequential algorithm.

The effect of speedup in $\sqrt{2}$ times can not be increased essentially 
by the same procedure if we increase a number of oracles involved in the 
simultaneous action. Nevertheless, a possibility of further speedup of the 
iterated quantum search still remains.

\section{Acknowledgements}

I am grateful to Victor Maslov for his support,
 to Oleg Khrustalev, Anatolyi Fomenko, Voislav Golo, Grigori Litvinov, 
Yuri Solomentsev and Kamil Valiev 
for the interesting discussions and help.

I am especially grateful to Lov Grover for his attention to my work and 
useful discussions and to Peter Hoyer for valuable comments to my 
previous works. I also thank Charles Bennett, David DiVincenzo, Lev Levitin, 
Richard Cleve, Richard Jozsa and others for the 
 fruitful discussions at the conference 
NASA QCQC'98.

I would like also to express my gratitude to Edward Farhi and Sam Gutmann who have read the first version of this work and suggested the simplification of the proof.

\appendix
\section{Appendix. An iterated approximation for solutions of differential 
equations}

\subsection{The method of simple iterations}

Here we present the first version of the proof.
Let $K,L,K_0$ be linear operators of the form ${\rm C}^m \ar{\rm C}^m$,
$K=K_0 +L$, $\sigma (t) :\ {\rm R} \ar {\rm C}^m$ be a vector function
with scalar argument.

Consider the following Cauchy problem for a system of linear differential 
equations with unknown vector function $x(t)$
\begin{equation}
\dot x(t)=Kx(t)+\sigma (t),\ \ x(0)=x_0 \in {\rm C}^m ,
\label{19}
\end{equation}
where $K_0$ is regarded as the main summand and $L$ determines a
perturbation such that a solution $x_0 (t)$ of the problem
$$
\dot x =K_0 x,\ \ x(0)=x_0
$$
is known. Then a solution of (21) can be obtained as a limit of the sequence 
$x_0 , x_1 ,\ldots $ of vector functions depending on 
$t$, which are determined by the following recursion
\begin{equation}
x_{i+1} =x_0 +\int\limits_{0}^{t} (Kx_i +\sigma )dt .
\label{20}
\end{equation}

Define the $i$-th difference between the approximations as
$\delta_i =x_i -x_{i-1} ,\ \ i=1,2,\ldots$. Then 
$\delta_1 = \int\limits_{0}^{t} Lx_0 dt ,$ $\delta_{i+1}
=\int\limits_{0}^{t} K\delta_i dt$.
Thus we conclude that
\begin{equation}
\delta_i =K^i \int\limits_{0}^{t} \int\limits_{0}^{t_1}
\int\limits_{0}^{t_2} \ldots\int\limits_{0}^{t_{i-2}} 
\delta_1 (t_{i-2} )dt_{i-2} \ldots dt_2 dt_1 dt .
\label{21}
\end{equation}

\subsection{The method of complex iterations}

The simple method from the previous section can not 
be immediately applied to our situation because of inconvenient form of our 
main matrix $M$. We apply the following idea.
In each step of approximation consider separately an approximations  on the 
group of $b,a,\a$ and then on $\b$. 
Here the main matrix $A=A_0 +B$ of the system for $b,a,\a$ can be represented as a 
result of small perturbation $B$ of almost nilpotent matrix which allows to 
estimate a deposit of perturbation.

These matrices with the notations of (16) have the form
\begin{equation}
A_0 =
\left(
\begin{array}{ccc}
0 &2 &4\\
-\frac{2}{N} &0 &2\\
0 &-\frac{2}{N} &0
\end{array}
\right)
, \ B=\left(
\begin{array}{ccc}
O_1 (\frac{1}{N})+O_{11}(\frac{1}{N^2}) &O_0 (\frac{1}{N})+O_{01}
(\frac{1}{N^2}) &O_{31}(\frac{1}{N})\\
O_{12}(\frac{1}{N^2}) &O_2 (\frac{1}{N})+O_{02}(\frac{1}{N^2})
&O_{32}(\frac{1}{N})\\
O_{13}(\frac{1}{N^2}) &O_{03}(\frac{1}{N^2}) &O_{33}(\frac{1}{N})
\end{array}
\right) .
\label{22}
\end{equation}
Consider the vectors 
$$
\begin{array}{cl}
|d_{\b} \rangle &=|-\frac{2}{N} +O{14} (\frac{1}{N^2}) , O_4
(\frac{1}{N})+O_{04}(\frac{1}{N^2}) , 2+O_{34}(\frac{1}{N}),
-2+O_{24}(\frac{1}{N})\rangle ,\\
|\g \rangle &=|-2+O_{21}(\frac{1}{N}), O_{22}(\frac{1}{N}),
O_{23}(\frac{1}{N})\rangle .
\end{array}
$$

Preparing to define by induction a sequence of approximations to the solution 
of (16), (17) we introduce the notations
$$
|\bar c_i \rangle =|b_i ,a_i ,\a_i \rangle ,\ 
|\tilde c_i \rangle = |\bar c_i ,\b_i \rangle .
$$

Then the Cauchy problem (16), (17) is equal to the system
\begin{equation}
\begin{array}{cl}
|\dot{\bar c} \rangle &=A|\bar c\rangle +\b |\g \rangle ,\\
\dot\b &=\langle d_{\b} |\tilde c\rangle
\end{array}
\label{23}
\end{equation}
with the initial condition (17).

Define the sequence of vector functions $|\tilde c_i \rangle$ 
 approximating the solution of (25), (17) by the following induction on $i$.

{\bf Basis}. $|\tilde c_0 \rangle =|\bar c_0 , \b_0 \rangle$,
where $|\bar c_0 \rangle =|b_0 ,a_0 ,\a_0 \rangle$ is a solution
of Cauchy problem $|\dot{\bar c}\rangle=A_0 |\bar c\rangle ,$
$\bar c(0)=|\frac{1}{N} ,\frac{1}{N} ,\frac{1}{N} \rangle$ given
by the approximations (18), $\b_0$ is a solution of equation
$\dot{\b}_0 =\langle d_{\b}|\tilde c_0 \rangle$ with the initial
condition $\b_0 (0)=\frac{1}{N}$.

{\bf Step}. $\bar c_{i+1}$ is a solution of Cauchy problem
$$
\dot{\bar c}_{i+1} =A\bar c_{i+1} +\b_i |\g_{\b} \rangle ,
\ \ \bar c_{i+1} (0)=|\frac{1}{N} ,\frac{1}{N} ,\frac{1}{N} \rangle ;
$$
$\b_{i+1}$ is the solution of $\dot\b_{i+1} =\langle d_{\b}|\tilde c_{i+1}
\rangle,\ \b_{i+1} (0)=\frac{1}{N}$ with the given 
vector function $\bar c_{i+1}$ obtained
above.

Put $\delta_i =\tilde c_i -\tilde c_{i-1} = |\delta_i^b ,\delta_i^a ,
\delta_i^{\a} ,\delta_i^{\b} \rangle ,\ i=1,2,\ldots$. The main fact 
concerning these approximations is the following

\newtheorem{Lemma}{Lemma}
\begin{Lemma}
For $0\leq t\leq\frac{\pi\sqrt{N}}{2\sqrt{2}}$ 
$$
|\delta_i^b |\leq\frac{10t^{i-1}}{N^{\frac{i}{2}}(i-1)!};\ 
|\delta^{\b} |,|\delta_i^a |\leq\frac{10t^{i-1}}{N^{\frac{i+1}{2}}(i-1)!};\ 
|\delta_i^{\a} |\leq\frac{10t^{i-1}}{N^{\frac{i}{2}+1}(i-1)!}.
$$
\end{Lemma}

This Lemma means that the vectors $\delta_i$ form summable row  with a module 
of sum less than $\frac{20}{\sqrt{N}}$ and thus we obtain Theorem.
To prove the key Lemma 1 we need to clarify a main property of matrix $A$.

 \subsection{A nilpotency of the main matrix}

Denote by $\e_i$ any number whose absolute value is less or equal to 
$(10/N)^j$. 
The main property of the matrix $A$ is represented by the following
Lemma.
\begin{Lemma}
For every $j=0,1,\ldots $
$$
A^{2j} =\left(
\begin{array}{ccc}
\e_j &\e_j &\e_{j-1}\\
\e_{j+1} &\e_j &\e_j\\
\e_{j+1} &\e_{j+1} &\e_j
\end{array}
\right) ,
\ A^{2j+1} =
\left(
\begin{array}{ccc}
\e_{j+1} &\e_j &\e_j\\
\e_{j+1} &\e_{j+1} &\e_j\\
\e_{j+1} &\e_{j+1} &\e_{j+1}
\end{array}
\right) .
$$
\end{Lemma}

{\it Proof}

Induction on $j$.

\subsection{Completion of the proof}

\ \ \ \ \ \ \ {\it Proof of Lemma 1}
\nn

Induction on $i$. 

{\bf Basis}. 

Consider the passage $\tilde c_0 \ar\tilde c_1$, consisting
 of two parts, we regard them as cases a) and b).

{\bf a)}. $\bar c_0 \ar\bar c_1$. This passage is the 
sequence
$\bar c_0 =\bar c_0^0 \ar\bar c_0^1 \ar\ldots \ar 
\bar c_0^t \ar\ldots \ar\bar c_1$, where 
$\bar c_0^{j+1} = \bar c(0)+\int\limits_{0}^t (A \bar c_0^j +
\b_0 \bar\g )dt$ ($\g$ in ordinary notations). The $j$-th difference will 
be $\Delta_j =
\bar c_0^{j} -\bar c_0^{j-1} =\int\limits_{0}^t A\Delta_{j-1} dt$,
$\Delta_1 =\int\limits_0^t (B\bar c_0^0 +\b_0 \bar\g )dt$.

Assume the following notation for an arbitrary function $F(t)$:
\newline 
${\int\limits_0^t}^{\{ k\}} F(t)dt=\int\limits_0^t \int\limits_
0^t \ldots \int\limits_0^t F(t)dt\ \ldots dt\ dt$.
We will write $\bar a\preceq \bar b$ iff an absolute value of each component 
of $\bar a$ does not exceed the absolute value of the corresponding component 
of $\bar b$.
Then there exists such numbers $l_s ,\ s=1,2,3$ that for every $k=1,2,\ldots$ 
we have 
${\int\limits_0^t}^{\{ k\}} \Delta_1 \preceq\frac{t^k}{k!}
|\frac{l_1}{\sqrt{N}} ,\frac{l_2}{N},\frac{l_3}{N^{3/2}} \rangle$.

Now applying sequentially the equality (23) and Lemma 1, we obtain:
$$
\Delta_j =A^{j-1} {\int\limits_0^t}^{\{ j-1\}} \Delta _1 dt 
\preceq \left(\frac{10t}{\sqrt{N}}\right) \frac{1}{(j-1)!}
\left|\frac{1}{\sqrt{N}} ,\frac{1}{N} ,\frac{1}{N^{3/2}} \right\rangle .
$$

Hence $\bar c_1 -\bar c_0 =\Delta_1 +\Delta_2 +\ldots \preceq\exp (
\frac{10t}{\sqrt{N}})|
\frac{1}{\sqrt{N}} ,\frac{1}{N} ,\frac{1}{N^{3/2}} \rangle .
$

{\bf b)}.
$\bar c_1 \ar \tilde c_1$.

Introduce the following scalar function $\Phi (b,a,\a )=\langle d_{\b}
|b,a,\a \rangle$. The difference for it is $\Delta\Phi_i =
\Phi_i -\Phi_{i-1}$.

The first $\b$-difference $\delta_1^{\b} =\b_1 -\b_0$ is the difference 
between the solutions of equations
$\dot\b_0 =\Phi (b_0 ,a_0 ,\a_0 )-2\b_0$ and
$\dot\b_1 =\Phi (b_1 ,a_1 ,\a_1 )-2\b_1$.

\begin{Lemma}
For each $i=1,2,\ldots$
$$
\delta_i^{\b} =\exp (-2t)\int\limits_0^t \exp (2t) \Delta\Phi_i dt .
$$
\end{Lemma}

{\it Proof}
\nn

$\delta_i^{\b} $ is the solution of the Cauchy problem 
$\dot\delta_i^{\b} +2\delta_i^{\b}=\Delta\Phi_i ,\ \ \delta_i^{\b} (0)=0$,
which can be obtained immediately. Lemma 3 is proved.

Applying Lemma 3 we obtain $|\delta_1^{\b}| \leq\frac{t}{2N\sqrt{N}}$.
Hence for the first step if $t\leq \sqrt{N}$ then
$\delta_1 \preceq |\frac{1}{\sqrt{N}},\frac{10}{N} ,\frac{10}{N^{3/2}} ,
\frac{10}{N^{3/2}}\rangle$.
Basis is complete.

{\bf Step}.

The passage $\tilde c_{i-1} \ar\tilde c_i$ consists of two
 sequential steps:

1) the passage from $\bar c_{i-1}$ to $\bar c_i$,

2) the passage from $\b {i-1} $ to $\b_i$.

Take up {\bf Step 1}. Applying the method of approximations (22)
 to the problem (25), we regard $\bar c_i$ as the limit of the sequence
$\bar c_i^0 ,\bar c_i^1 ,\ldots$, where $\bar c_i^0 =\bar c_{i-1}$,
$$
\bar c_i^j =\bar c_i^0 +\int\limits_0^t (A\bar c_i^{j-1} +\b _{i-1} 
\bar \g )dt.
$$
The accuracy of $j$-th approximation relative to the previous one 
is determined by the difference $\Delta^j =\bar c_i^j -\bar c_i^{j-1} 
=|\Delta_j^b ,\Delta_j^a ,\Delta_j^{\a} \rangle$,
given by 
\begin{equation}
\begin{array}{cl}
\bar\Delta^j &=A^{j-1}{\int\limits_0^t}^{\{ j-1\}} 
\bar\Delta^1 (t)dt,\ \ j>1,\\
\bar\Delta^1 &=\int\limits_0^1 \delta_{i-1}^{\b} \bar\g dt.
\end{array}
\label{25}
\end{equation}

Taking into account inductive hypothesis we obtain for the first difference
\newline
$\bar \Delta_i^1 \preceq \int\limits_0^t \frac{t^{i-1}}{(i-1)! N^{i/2}}
|\frac{10}{N} ,\frac{10}{N^2} ,\frac{10}{N^2} \rangle dt$. For the next 
differences we have \newline
${\int\limits_0^t}^{\{ j-1\} } \bar\Delta_i^1 dt \preceq
\frac{t^{i-1}}{(i-1)! N^{i/2}} \left|\frac{3t^{j-1}}{(j-1)!N} ,
\frac{10t^{j-1}}{(j-1)!N^2} , \frac{10t^{j-1}}{(j-1)!N^2} \right\rangle$.

Calculating the differences accordingly to the formula (26) and applying
Lemma 2 we obtain
$$
\begin{array}{cl}
\bar\Delta_i^{2j} &=\frac{t^{i-1}}{(i-1)! N^{i/2}}
\left|\frac{30t^{2j-1} \e_j}{(2j-1)! N}, \frac{10t^{2j-1} 
\e_{j-1}}{(2j-1)!N^2} ,
\frac{10t^{2j-1} \e_{j-1}}{(2j-1)! N^2} \right\rangle ,\\
\bar\Delta_i^{2j+1} &=\frac{t^{i-1}}{(i-1)! N^{i/2}}
\left|\frac{30t^{2j} 10^j}{(2j+1)! N^{j+1}} , \frac{30t^{2j} 10^j}{(2j+1)! 
N^{j+2}} ,
\frac{30t^{2j} 10^j}{(2j+1)! N^{j+2}}\right\rangle .
\end{array}
$$
Summing over all $j$ gives the inequalities for $\delta^b,\delta^a ,
\delta^{\a}$ from the statement of Lemma 1.
 As for $\delta^{\b}$, we estimate it as in the basis of induction,
Then Lemma 3 gives $|\delta_i^{\b} |\leq \frac{t^i}{NN^{i/2} i!}$.
Step of induction is complete. Lemma 1 is proved. Theorem is proved.
\end{document}